\documentclass{snmp2003}

\begin{document}

\FirstPageHeading{A.J. Nurmagambetov}

\ShortArticleName{The Sigma-Model Representation for the Duality-Symmetric
D=11 Supergravity} 

\ArticleName{The Sigma-Model Representation for the\\
Duality-Symmetric D=11 Supergravity}

\Author{Alexei J. NURMAGAMBETOV~$^\dag$ }
\AuthorNameForHeading{A.J. Nurmagambetov}
\AuthorNameForContents{NURMAGAMBETOV A.J.}
\ArticleNameForContents{The Sigma-Model Representation for the
Duality-Symmetric D=11 Supergravity}

\Address{$^\dag$~Institute for Theoretical Physics, NSC ``Kharkov
Institute of Physics and Technology"\\1 Akademicheskaya Str.,
61108, Kharkov, Ukraine} \EmailD{ajn@kipt.kharkov.ua}


\Abstract{In this paper which is based on the results obtained in
collaboration with Igor Bandos and Dmitri Sorokin I present the
extention of the ``doubled fields" formalism by Cremmer, Julia,
L\"u and Pope to the supersymmetric case and lifting this
construction onto the level of the proper duality-symmetric action
for D=11 supergravity. Further extension of the doubled field
formulation to include dynamical sources is also discussed.}

\section{Introduction}

This year we can truly commemorate the 25th anniversary from the
day when eleven-dimensional supergravity
\cite{Nurmagambetov:cremmer&julia&scherk} was born. It is worth
mentioning the importance of this event since supergravity in D=11
space-time dimensions is a kind of special theory possessing a
number of non-trivial features. First of all, this theory is the
maximally-dimensional supersymmetric theory of gravity interacting
with antisymmetric tensor fields for the standard signature of
metric with one time-like direction. Second, the theory has
relatively simple structure of the action in compare with other
low-dimensional maximal supergravities. The third point consists
in the observation that the massless spectrum of a theory derived
upon a compactification from D=11 down to D=4 does not contain
particles with spin higher than two that is an indication of
having a consistent four-dimensional interacting theory. Last but
not least the eleven-dimensional supergravity is the low-energy
limit of M-theory unifying non-perturbatively five different
theories of superstrings. Since not much has been known about
M-theory, D=11 supergravity is very useful tool to investigate the
M-theory structure.

One of the ways to do so is to consider different routes of
dimensional reduction to establish the connection between
different low-energy effective actions for superstrings which are
nothing but supergravities and to find a relevant scheme that
would be in accordance with four-dimensional phenomenology.
Following this way another notion, the duality, becomes important
and allows us to establish the symmetry structure in reduced
theories which is originally hidden. Moreover, the duality gives a
fresh look on the symmetry structure in higher-dimensional
supergravities that allows one to find new way for describing in a
uniform duality invariant manner gauge and internal symmetries of
these theories.

It should be emphasized that the notion of duality in the context
of searching new symmetries in dimensionally reduced
supergravities \cite{Nurmagambetov:cremmer&julia} appeared one
year after discovering the eleven-dimensional supergravity. There
discovered was a hidden symmetry structure of maximal
low-dimensional supergravities that extends the naively expected
global symmetry groups arising in the process of dimensional
reduction on multidimensional tori. Some important observations
were made that was crucial to develop the technique to derive
later on the low-dimensional theories of supergravities by
reduction on spheres and manifolds with non-trivial holonomy
groups.

The aim of this paper is twofold. First, we would like to review
the old results reminding ideas and methods that led to the
discovery of hidden global symmetry structure in maximal
supergravities and to trace how these ideas and methods naturally
get transformed for the description of local symmetries in such
theories \cite{Nurmagambetov:cremmer&julia&lu&pope}. And second,
we present new results
\cite{Nurmagambetov:bandos&nurmagambetov&sorokin} which generalize
the recent construction of local symmetries in uniform duality
invariant manner. Though we shall note only the key points that
are necessary for the discussion in what follows, all the details
can be read off original papers
\cite{Nurmagambetov:cremmer&julia}, \cite{Nurmagambetov:julia98},
\cite{Nurmagambetov:cremmer&julia&lu&pope},
\cite{Nurmagambetov:julia00}, \cite{Nurmagambetov:cremmer81},
\cite{Nurmagambetov:bandos&nurmagambetov&sorokin}.

\section{Hidden global symmetries in maximal supergravities}

Let me get started with recalling that $D$-dimensional maximal
supergravity is the theory which follows from D=11 supergravity by
dimensional reduction on torus $T^{(11-D)}$. Typically, such a
theory is expected to be invariant under the product of a
non-compact global group and a compact local group
$SL(11-D,R)_{Global}\otimes SO(11-D)_{Local}$. The former group
acts on tensors and scalars, while the latter one acts on fermions
and scalars. However, close inspection reveals the hidden symmetry
group structure extending the structure above. This is due to
``conspiracy" of scalars and tensors coming in the reduction
process from gravitational and tensor gauge fields sectors and of
scalars and tensors appearing after dualizing a higher $p$-rank
tensor field with $p\ge D/2$ into a $(D-2-p)$-rank tensor field.
Recasting the fields the following symmetry structure can be
recovered \cite{Nurmagambetov:cremmer81}
\begin{center}
\begin{tabular}{|c|c|c|}
\hline\hline $D$&$G_{Global}$&$H_{Local}$\\ \hline $D=11$&$E_0\sim
{\bf 1}$&${\bf 1}$\\ \hline $D=10$&$E_1\sim SO(1,1)/Z_2$&${\bf
1}$\\ \hline $D=9$&$E_2=GL(2,R)$&$SO(2)$\\ \hline
$D=8$&$E_{3(+3)}=SL(3,R)\otimes SL(2,R)$&$U(2)=SO(3)\otimes
SO(2)$\\ \hline $D=7$&$E_{4(+4)}=SL(5,R)$&$Usp(4)=SO(5)$\\ \hline
$D=6$&$E_{5(+5)}=SO(5,5)$&$Usp(4)\otimes Usp(4)$\\ \hline
$D=5$&$E_{6(+6)}$&$Usp(8)$\\ \hline $D=4$&$E_{7(+7)}$&$SU(8)$\\
\hline $D=3$&$E_{8(+8)}$&$SO(16)$\\ \hline \hline
\end{tabular}\end{center}

Let me now introduce the basic notion of a twisted self-duality
condition which will be important in what follows. To this end
note that when the dimension $D$ is odd the $E_{11-D}$ symmetry is
realised on the gauge potentials and is an invariance of the
Lagrangian. However, as soon as the space-time dimension is even,
the $E_{11-D}$ symmetry for field strengths of degree $D/2$ is
realised on the field strengths rather than on the potentials and
is an invariance of the equations of motion and Bianchi identities
only. But one can still realise such a symmetry at the level of
equations of motion on the potentials if one introduces their
additional dual partners. After that the equations of motion take
the form of the twisted self-duality conditions
\cite{Nurmagambetov:cremmer&julia}
\begin{equation}\label{Nurmagambetov:eq1}
{\cal V}{\cal F}^{(D/2)}=\Omega {\cal V}\ast{\cal F}^{(D/2)},
\end{equation}
where $\Omega$ is the symplectic metric of the $Sp$ group that
contains the $E_{11-D}$ as a subgroup, ${\cal V}$ is the matrix of
scalars in the $SL$ subgroup of the $E_{11-D}$ and $\ast$ stands
for $D$-dimensional Hodge operator.

The meaning of the twisted self-duality conditions is clear. As
soon as we introduced additional dual potentials, the number of
degrees of freedom for corresponding tensor fields became twice
higher than we need. The twisted self-duality condition reduces
this number turning back to the standard situation. However,
curiously, the corresponding equations of motion for our theory
have been encoded in these conditions.

What we have learned from these observations? We should emphasize
the role of duality in the establishing the global hidden
symmetries in maximal supergravities. Later on dualities were
recognized as a basic tool to establish hidden non-perturbative
symmetries of M-theory. The next point is that we have another
description for the dynamics of self-dual fields having the field
strengths of the degree of $D/2$. Two questions arise: Could we
apply similar approach to describe the dynamics of fields dual to
each other with respect to the Hodge operation and which are not
necessary self-dual? And could we generalize the construction to
describe local symmetries of the theory?

The answer came from the observation made by Cremmer, Julia, L\"u
and Pope \cite{Nurmagambetov:cremmer&julia&lu&pope}. Motivated by
studying the global symmetry groups in maximal supergravities in
the context of the M-theory conjecture, the authors observed that
doubling the gauge fields, as well as dilatonic scalars and
axionic scalars by their duals one can present the bosonic
equations of motion for this sector of supergravities, starting
from D=11 supergravity, in the form which generalizes the twisted
self-duality conditions
\begin{equation}\label{Nurmagambetov:eq2}
\ast{\cal G}={\cal S}{\cal G}.
\end{equation}
Here ${\cal G}$ is a combined single field written in terms of the
exponential of linear combinations of generators for each field
and its double, with all the potentials as coefficients. ${\cal
S}$ is a pseudo-involution that exchanges the generators for
fields and those for their partners under the doubling. Complete
construction is a remnant of $G/H$ coset structure of scalar
fields well known in supergravities and can be regarded as its
generalization. Since the dynamics of scalars is described by the
$G/H$ sigma-model action one can expect the sigma-model structure
for ``doubled fields action" constructed out of ${\cal G}$, if the
action does exist. To find such an action, let me to be more
precise confining myself to the space-time dimension eleven and
presenting the details of the construction by Cremmer, Julia, L\"u
and Pope for D=11 supergravity. Although this case is simple in
comparison with even ten-dimensional type IIA supergravity
obtained from D=11 supergravity by dimensional reduction, it is
generic and possesses all features for the doubled field approach.

\section{Doubled field formulation of D=11 supergravity}

From the Lagrangian for the bosonic sector of D=11 supergravity
\cite{Nurmagambetov:cremmer&julia&scherk}
\begin{equation}\label{Nurmagambetov:eq3}
{\cal L}=-\sqrt{g}~R-{1\over 48}\sqrt{g}~F_{mnpq}F^{mnpq}-{1\over
3}A^{(3)}\wedge F^{(4)}\wedge F^{(4)},\qquad F^{(4)}=dA^{(3)}
\end{equation}
we get the second order equation of motion for the $A^{(3)}$ gauge
field
\begin{equation}\label{Nurmagambetov:eq4}
d(\ast F^{(4)}-A^{(3)}\wedge F^{(4)})=0
\end{equation}
that can be presented as the Bianchi identity for the dual field
$A^{(6)}$
\begin{equation}\label{Nurmagambetov:eq5}
dF^{(7)}=0,\qquad F^{(7)}=dA^{(6)}+A^{(3)}\wedge F^{(4)}.
\end{equation}
Now forget for a while the dynamical origin of
$F^{(7)}=dA^{(6)}+A^{(3)}\wedge F^{(4)}$ and introduce it as an
independent partner of $F^{(4)}$. These field strengths are
invariant under the local gauge transformations
\begin{equation}\label{Nurmagambetov:eq6}
\delta A^{(3)}=\Lambda^{(3)},\qquad \delta
A^{(6)}=\Lambda^{(6)}-\Lambda^{(3)}\wedge A^{(3)}
\end{equation}
with closed forms $\Lambda^{(3)}$, $\Lambda^{(6)}$ associated with
the so-called large gauge transformations
\cite{Nurmagambetov:llps}. Because of the presence of ``bare"
$A^{(3)}$ in $F^{(7)}$ and therefore in $\delta A^{(6)}$ that is
traced back to the presence of the Chern-Simons term in the
Lagrangian (\ref{Nurmagambetov:eq3}) the large gauge
transformations are non-abelian
\begin{equation}\label{Nurmagambetov:eq7}
[\delta_{\Lambda^{(3)}_1},\delta_{\Lambda^{(3)}_2}]=\delta_{\Lambda^{(6)}},
\qquad [\delta_{\Lambda^{(3)}},\delta_{\Lambda^{(6)}}]=
[\delta_{\Lambda^{(6)}_1},\delta_{\Lambda^{(6)}_2}]=0.
\end{equation}
These relations can be associated with a superalgebra generated by
a ``Grassmann-odd" generator $t_3$ and a commuting generator $t_6$
\begin{equation}\label{Nurmagambetov:eq8}
\{t_3,t_3\}=-2t_6,\qquad [t_3,t_6]=[t_6,t_6]=0
\end{equation}
after that one can realize an element of the supergroup
\begin{equation}\label{Nurmagambetov:eq9}
{\cal A}=e^{t_3 A^{(3)}} e^{t_6 A^{(6)}}
\end{equation}
and introduce the Cartan form
\begin{equation}\label{Nurmagambetov:eq10}
{\cal G}=d{\cal A}{\cal A}^{-1}=F^{(4)}t_3+F^{(7)}t_6,
\end{equation}
which by definition satisfies the Maurer-Cartan equation called
sometimes the zero-curvature condition
\begin{equation}\label{Nurmagambetov:eq11}
d{\cal G}+{\cal G}\wedge {\cal G}=0.
\end{equation}
To impose the duality relation between a priori independent field
strengths and arriving therefore at the standard number of degrees
of freedom one introduces the pseudo-involution operator ${\cal
S}$ which interchanges the generators $t_3$ and $t_6$
\begin{equation}\label{Nurmagambetov:eq12}
{\cal S}t_3=t_6,\qquad {\cal S}t_6=t_3,\qquad {\cal S}^2=1.
\end{equation}
Using ${\cal S}$ and the Hodge operator one can immediately check
that the following condition
\begin{equation}\label{Nurmagambetov:eq13}
\ast{\cal G}={\cal S}{\cal G}
\end{equation}
reproduces correctly the duality relations between the field
strenghts and therefore reduces tensors' degrees of freedom to the
correct number. Moreover, when this condition holds the
Maurer-Cartan equation amounts to second order equations of motion
for $F^{(4)}$ and $F^{(7)}$.

Since the relation (\ref{Nurmagambetov:eq13}) enjoys all necessary
properties of the condition (\ref{Nurmagambetov:eq1}) it also
shares the name of the twisted self-duality condition.

So, we have explicitly described the doubled field approach to the
bosonic sector of D=11 supergravity. However, we should still add
the fermions and construct a proper action for the doubled fields.
We are now turning to these points.

\section{Generalization of doubled field approach to fermions and
lifting it onto the level of action}

Adding the fermions is a simple task since what one should do is
to extend ${\cal G}$ with the superalgebra valued element ${\cal
C}=-C^{(4)}t_3+C^{(7)}t_6$
\cite{Nurmagambetov:bandos&nurmagambetov&sorokin}
\begin{equation}\label{Nurmagambetov:eq14}
{\cal G}\longrightarrow {\cal G}+{\cal C},
\end{equation}
where $C^{(4)}$ and $C^{(7)}$ are defined as
\begin{equation}\label{Nurmagambetov:eq15}
C^{(4)}=-{1\over 4}\bar{\psi}\wedge \Gamma^{(2)}\wedge \psi,~
C^{(7)}={i\over 4}\bar{\psi}\wedge \Gamma^{(5)}\wedge \psi,~
\Gamma^{(n)}={1\over n!}dx^{m_n}\wedge\dots\wedge
dx^{m_1}\Gamma^{(n)}_{m_1\dots m_n},
\end{equation}
and $\psi^{\alpha}$ is the one-form associated with the Majorana
spin 3/2 gravitino $\psi^{\alpha}_m$.

After this replacement the twisted self-duality condition becomes
\begin{equation}\label{Nurmagambetov:eq16}
\ast ({\cal G}+{\cal C})={\cal S}({\cal G}+{\cal C})
\longrightarrow ({\cal S}-\ast)({\cal G}+{\cal C})=0.
\end{equation}
Therefore, if one finds a way to construct the action from which
the twisted self-duality condition (\ref{Nurmagambetov:eq16}) will
follow as an equation of motion the task will be completed.
Fortunately, this way exists and is based on the PST technique
\cite{Nurmagambetov:pasti&sorokin&tonin} developed to construct
the Lagrangians in theories with self-dual or duality-symmetric
fields.

Applying the PST technique the twisted self-duality condition is
reproduced from the following action
\cite{Nurmagambetov:bandos&nurmagambetov&sorokin}
\begin{gather}
S=S_{EH}+S_{\psi}-Tr\int_{{\cal M}^{11}}\, {1\over 2}({\cal
G}+{1\over 2}{\cal C})\wedge ({\cal S}-\ast){\cal C}\nonumber\\
-Tr\int_{{\cal M}^{11}}\,\left[{1\over 4}\ast{\cal G}\wedge {\cal
G}-{1\over 12}{\cal G}\wedge {\cal S}{\cal G}-{1\over 4}\ast
i_v({\cal S}-\ast)({\cal G}+{\cal C})\wedge i_v({\cal
S}-\ast)({\cal G}+{\cal C})\right],\label{Nurmagambetov:eq17}
\end{gather}
where $S_{EH}$ and $S_{\psi}$ stand for the Einstein-Hilbert
action and for the fermionic kinetic term
\cite{Nurmagambetov:cremmer&julia&scherk},
\begin{equation}\label{Nurmagambetov:eq18}
v={d a(x)\over \sqrt{-(\partial a)^2}}
\end{equation}
is the one-form constructed out the PST scalar auxiliary field
(cf. \cite{Nurmagambetov:pasti&sorokin&tonin}) which ensures the
covariance of the action and
\begin{equation}\label{Nurmagambetov:eq19}
Tr(t_3 t_3)=-Tr (t_6 t_6)=-1,\qquad Tr(t_3 t_6)=0.
\end{equation}
As usual we have denoted by $i_v$ the inner product of the vector
field $v_m$ with a form.

It is worth mentioning that the action just presented is the
generalization of the sigma-model action having typically the form
\begin{equation}\label{Nurmagambetov:eq20}
S \propto \int_{{\cal M}^D}\, Tr(\ast{\cal G}\wedge {\cal G})
\end{equation}
and therefore we have derived the result which we expected from
the beginning.

It is an instructive exercise to rewrite the sigma-model action to
the standard for D=11 supergravity form. After some manipulations
with taking into account the definitions of ${\cal G}$ and ${\cal
S}$ one can arrive at
\begin{equation}\label{Nurmagambetov:eq21}
S=S_{CJS}+\int_{{\cal M}^{11}}\, {1\over 2}i_v{\cal
F}^{(4)}\wedge\ast i_v{\cal F}^{(4)}
\end{equation}
with $S_{CJS}$ being the standard action by Cremmer, Julia and
Scherk \cite{Nurmagambetov:cremmer&julia&scherk} and ${\cal
F}^{(4)}=F^{(4)}-\ast F^{(7)}$. This is the action for the
duality-symmetic D=11 supergravity
\cite{Nurmagambetov:bandos&berkovits&sorokin} from which one can
dynamically derive the duality condition ${\cal F}^{(4)}=0$. Note
also that this action depends on both $A^{(3)}$ and $A^{(6)}$
potentials therefore the name duality-symmetric, and on the shell
of the duality condition ${\cal F}^{(4)}=0$ it coincides with the
Cremmer-Julia-Scherk action.

\section{The sigma-model representation of D=11 supergravity coupling
to the M-sources}

It is well known that a M2 and a M5 branes are dynamical sources
for $A^{(3)}$ and $A^{(6)}$ D=11 supergravity gauge fields
\cite{Nurmagambetov:bandos&berkovits&sorokin},
\cite{Nurmagambetov:alwis}, \cite{Nurmagambetov:sorokin} and
therefore one may wonder what is the way of extension of the
doubled fields formalism to describe dynamical coupling of D=11
supergravity to the M-branes' sources? To get the answer let me
consider the case of coupling of D=11 supergravity to single
closed M2 and M5 branes which is naively described by the
following set of Bianchi identities and equations of motion (cf.
\cite{Nurmagambetov:alwis},
\cite{Nurmagambetov:bandos&berkovits&sorokin},
\cite{Nurmagambetov:lmt}, \cite{Nurmagambetov:lm})
\begin{gather}
d\hat{F}^{(4)}=\ast J^{(6)}_m,\nonumber\\
d\ast
\hat{F}^{(4)}=\hat{F}^{(4)}\wedge\hat{F}^{(4)}-2h^{(3)}\wedge\ast
J^{(6)}_m+\ast J^{(3)}_e,\nonumber\\
h^{(3)}=\star h^{(3)}+n.~l.~t.,\label{Nurmagambetov:M5em}
\end{gather}
where
\begin{gather}
\hat{F}^{(4)}=dA^{(3)}+\ast G^{(7)}_m,\qquad d\ast G^{(7)}_m=\ast
J^{(6)}_m,\qquad d\ast G^{(4)}=\ast J^{(3)}_e,\nonumber
\end{gather}
and
\begin{gather}
h^{(3)}=db^{(2)}-A^{(3)}\nonumber
\end{gather}
is the gauge invariant M5 worldvolume field strength constructing
out of the self-dual worldvolume gauge field $b^{(2)}$ and the
pullback of the target-space gauge field $A^{(3)}$. The last
equation of (\ref{Nurmagambetov:M5em}) that generalizes ordinary
self-duality condition contains non-linear corrections coming from
the Dirac-Born-Infeld-like structure of the M5 kinetic part (see
\cite{Nurmagambetov:schwarz}, \cite{Nurmagambetov:pst},
\cite{Nurmagambetov:blnpst}, \cite{Nurmagambetov:apps},
\cite{Nurmagambetov:blnpst1} for details) and $J_{e,m}$ are the
electric and magnetic currents due to the M2 and M5 branes. To
distinguish the Hodge operation in target space and on the
worldvolume of branes we have denoted the latter by $\star$.

It is easy to verify that the equation of motion of $A^{(3)}$
gauge field leads to the following expression for the field
strength dual to $\hat{F}^{(4)}$
\begin{equation}\label{Nurmagambetov:F7em}
\hat{F}^{(7)}=dA^{(6)}+A^{(3)}\wedge dA^{(3)}-2h^{(3)}\wedge \ast
G^{(7)}_m+\ast G^{(4)}_e,
\end{equation}
if one admits regularization for ill-defined product of two
delta-functions $\ast G^{(7)}_m\wedge \ast G^{(7)}_m$ entering the
$A^{(3)}$ equation of motion with $\ast G^{(7)}_m\wedge \ast
G^{(7)}_m=0$ \cite{Nurmagambetov:dght}, \cite{Nurmagambetov:bg}.
Such a regularization has been used in
\cite{Nurmagambetov:bandos&berkovits&sorokin} and is good enough
as far as the presence of classical gravitational anomaly due to
the M5-brane is neglected that we will suppose for the sake of
simplicity. However, it should be emphasized that the appearance
of product of delta-functions is ``a symptom of attempting to
treat in classical supergravity what really should be treated in
quantum M-theory" \cite{Nurmagambetov:hw}. This problem has been
approached and solved in a rigorous way using a notion of
distributions in \cite{Nurmagambetov:lmt},
\cite{Nurmagambetov:lm}.

The relevant part of the action which leads to the set of
equations (\ref{Nurmagambetov:M5em}) has the following generating
for the doubled field sigma-model action form
\begin{gather}
{\cal L}^{(11)}={1\over 4}\hat{F}^{(4)}\ast\hat{F}^{(4)}+{1\over
4}i_v\hat{\cal F}^{(4)}\ast i_v\hat{\cal F}^{(4)}-{1\over
4}\hat{F}^{(7)}\ast\hat{F}^{(7)}-{1\over 4}i_v\hat{\cal
F}^{(7)}\ast i_v\hat{\cal F}^{(7)}+{1\over
6}F^{(4)}F^{(7)}\nonumber\\
+[{\cal L}^{(6)}_{M5~kin.}-{1\over 2}
(A^{(6)}+db^{(2)}A^{(3)})]\ast J^{(6)}_m + [{\cal
L}^{(3)}_{M2~kin.}-{1\over 2} A^{(3)}]\ast J^{(3)}_e\nonumber\\
-{1\over 2}h^{(3)}\hat{F}^{(4)}\ast
G^{(7)}_m.\label{Nurmagambetov:dsL11em}
\end{gather}

To present the latter as the sigma-model-like action one should
extend the Cartan form ${\cal G}$ with the superalgebra valued
element ${\bf g}=t_6 (\ast G^{(4)}_e-2h^{(3)}\ast
G^{(7)}_m)+t_3\ast G^{(7)}_m$, i.e.
\begin{gather}
{\cal G}=d{\cal A}{\cal A}^{-1} \rightarrow {\cal G}=d{\cal
A}{\cal A}^{-1}+{\bf g}\nonumber
\end{gather}
after that the twisted self-duality condition becomes
\begin{equation}\label{Nurmagambetov:twe}
\ast {\cal G}={\cal S}{\cal G},
\end{equation}
but the zero-curvature condition is realized now as
\begin{equation}\label{Nurmagambetov:zcce}
d\tilde{\cal G}+\tilde{\cal G}\wedge \tilde{\cal G}=0,\qquad
\tilde{\cal G}={\cal G}-{\bf g}.
\end{equation}
When the twisted self-duality condition holds, i.e. when
\begin{equation}\label{Nurmagambetov:twer}
\ast \hat{F}^{(4)}=\hat{F}^{(7)},\qquad
\ast\hat{F}^{(7)}=\hat{F}^{(4)},
\end{equation}
equations that follow from the zero-curvature condition
(\ref{Nurmagambetov:zcce})
\begin{gather}
dF^{(4)}=0,\qquad dF^{(7)}=F^{(4)}\wedge F^{(4)}\nonumber
\end{gather}
are the same second order equations of motion that follow from the
action (\ref{Nurmagambetov:dsL11em}).

Therefore, the sigma-model doubled field action describing the
coupling of closed M2 and M5 branes to D=11 supergravity has the
following form
\begin{gather}
S=S_{EH}-Tr\int_{{\cal M}^{11}}\,\{ {1\over 4}\ast{\cal G}\wedge
{\cal G} -{1\over 12}\tilde{\cal G}\wedge {\cal S}\tilde{\cal G}
-{1\over 4}\ast i_v({\cal S}-\ast){\cal G}\wedge i_v({\cal
S}-\ast){\cal G} \}\nonumber\\
+\int_{{\cal M}^3}\, [ e^a\wedge\star e^b \eta_{ab}-{1\over
2}A^{(3)}]+\int_{{\cal M}^6}\, {\cal L}^{(6)}_{M5~kin.}-{1\over
2}\int_{{\cal M}^6}\,[A^{(6)}+db^{(2)}\wedge A^{(3)}]\nonumber\\
-{1\over 2}\int_{{\cal M}^{11}}\, h^{(3)}\wedge
\hat{F}^{(4)}\wedge\ast G^{(7)}_m.\label{Nurmagambetov:smem}
\end{gather}

\section{Discussion}

In conclusion let me summarize the obtained results. Following the
doubled field approach by Cremmer, Julia, L\"u and Pope we have
filled the gap in extending this construction to add fermions and
have formulated the supersymmeric action in such a way that the
twisted self-duality condition which encodes the dynamics of the
gauge fields follows from the action as an equation of motion. As
the next step towards a fully fledged theory we have extended the
construction to include the dynamical sources for antisymmetric
gauge fields and have presented the action describing the system
of D=11 supergravity interacting with dynamical branes.

The structure of the doubled field action is generic for all
maximal supergravities although the relevant superalgebra becomes
much more complicated (cf.
\cite{Nurmagambetov:cremmer&julia&lu&pope}). The case of D=10 type
IIB supergravity which can not be recovered by dimensional
reduction from D=11 supergravity also admits the relevant
superalgebra structure. This fact motives understanding the
T-duality between type IIA and type IIB theories in frames of the
doubled field approach. In our formulation we left the gravity
sector out of consideration due to the absence of clear
understanding how to present that in a duality-symmetric manner.
Having a recipe it will be interesting to realise the sigma-model
construction for the complete sector of fields of D=11
supergravity that possibly could give a new insight on the
mysterious M-theory structure.

\subsection*{Acknowledgements}

I am very grateful to Igor Bandos and Dmitri Sorokin for pleasant
collaboration and for very fruitful and illuminating discussions.
Special thanks to Xavier Bekaert for his suggestion to extend the
construction to include the sources and for very pleasant
discussion. This work is supported in part by the Grant N
F7/336-2001 of the Ukrainian State Fund for Fundamental Research
and by the INTAS Research Project N 2000-254.

\LastPageEnding

\end{document}